\documentclass[twocolumn,prd,nofootinbib,superscriptaddress,tightenlines,8pt]{revtex4}
\pdfoutput=1
\usepackage{fancyhdr}
\usepackage{amsmath,amsfonts,amssymb}

\usepackage{color}
\definecolor{red}{rgb}{1,0,0}
\definecolor{gre}{rgb}{0,0.6,0}
\definecolor{blu}{rgb}{0,0,1}

\usepackage[pdftex]{graphicx}
\usepackage{mathrsfs}
\DeclareGraphicsExtensions{.jpg, .JPG , .png , .pdf, .gif}
\def\be{\begin{equation}}
\def\ee{\end{equation}}

\renewcommand{\delta}{a}

\begin{document}

\title{Phenomenology of bouncing black holes in quantum gravity:\\
 a closer look 
 }

\date{\today}

\author{Aur\'elien Barrau}
\email{aurelien.barrau@cern.ch}
\affiliation{
Laboratoire de Physique Subatomique et de Cosmologie, Universit\'e Grenoble-Alpes, CNRS-IN2P3\\
53,avenue des Martyrs, 38026 Grenoble cedex, France\\
}%

\author{Boris Bolliet}
\email{boris.bolliet@ens-lyon.fr}
\affiliation{
Laboratoire de Physique Subatomique et de Cosmologie, Universit\'e Grenoble-Alpes, CNRS-IN2P3\\
53,avenue des Martyrs, 38026 Grenoble cedex, France\\
}%

\author{Francesca Vidotto}
\email{fvidotto@science.ru.nl}
\affiliation{
Radboud University, Institute for Mathematics, Astrophysics and Particle Physics\\
Mailbox 79, P.O. Box 9010, 6500 GL Nijmegen, The Netherlands
}%

\author{Celine Weimer}
\email{celinew@kth.se}
\affiliation{
Laboratoire de Physique Subatomique et de Cosmologie, Universit\'e Grenoble-Alpes, CNRS-IN2P3\\
53,avenue des Martyrs, 38026 Grenoble cedex, France\\
}%

\begin{abstract}
\vskip1em
It was recently shown that black holes could be bouncing stars as a consequence of quantum gravity. We investigate the astrophysical signals implied by this hypothesis, focusing on primordial black holes. We consider different possible bounce times and study the integrated diffuse emission.\\
\vskip4em
\end{abstract}

\maketitle


\section{The model}
A new possible window for observing quantum gravitational effects has been recently pointed out 
in \cite{Rovelli2014} (some details were refined in \cite{Lorenzo}). The idea is grounded on a result of loop cosmology \cite{Ashtekar2006}: when matter or radiation reaches the Planck density, quantum gravity generates a sufficient pressure to counterbalance the classically attractive gravitational force. In a black hole, matter's collapse could stop before the central singularity is formed. The standard event horizon of the black hole can be replaced by an apparent horizon \cite{Ashtekar:2005cj} which is locally equivalent to an event horizon, but from which matter can eventually bounce out. The model is not specific to loop quantum gravity (for instance a similar scenario can be realized in asymptotic safety \cite{Saueressig:2015qy}). 
The case of non-singular black holes has been investigated by many authors 
\cite{Bardeen1968non,
Frolov:1981mz,
Roman:1983zza,
Casadio:1998yr,
AyonBeato:2000zs,
Mazur:2001fv, 
dymnikova2002cosmological,
hayward2006formation,
ModestoLQGBH,
Visser:2009pw,
Modesto:2010rv,
carr2011,
LitimASBH,
BambiModestoKerr,
Bambi:2013caa,
Frolov:BHclosed, 
Mersini-Houghton:2014yq,
ModestoNonLocalBH,
Frolov:2015bta}.

A heuristic description of the model we are studying can be given as follows. When the density of matter becomes high enough, quantum gravity effects generate sufficient pressure to compensate the matter's weight, the collapse ends, and matter bounces out. A collapsing ``black hole" might avoid sinking into the $r=0$ singularity, as much as an electron in a Coulomb potential does not sink all the way into $r = 0$ because of quantum mechanical effects. The picture is close to Giddings's remnant scenario \cite{giddings} but with a macroscopic remnant developing into a white hole.

The phenomenology associated with this scenario was considered in  \cite{Barrau2014}, opening the fascinating possibility to detect quantum gravity effects far below the Planck energy. It was shown there that primordial black holes (PBHs) could generate a signal in the 100 MeV range, possibly compatible with very fast gamma-ray bursts \cite{nakar}. Observability is made possible by the amplification due to the large ratio of the black hole lifetime over the Planck time \cite{Amelino:13}.\\

The scenario was developed in \cite{Rovelli2014-2} with the discovery of an explicit metric satisfying Einstein's equations everywhere outside the quantum region. The model describes a quantum tunneling from a classical in-falling black hole to a classical emerging white hole. The process is seen in extreme ``slow motion'' from the outside because of the huge time dilatation inside the gravitational potential: this is why massive black holes would appear to us as long living black holes. Only light black holes --as primordial black holes-- are expected to yield observational signatures of this model because the time required for the bounce to occur can then be smaller that the current age of the Universe. 

Outside the horizon, the quantum effects are small at any time but their time integration can lead to important cumulative effects leading to a dramatic revision of the usual scenario. After a sufficiently long time, the black hole can tunnel to a white hole. This phenomenon is similar to the cosmological bounce studied in loop quantum cosmology \cite{Ashtekar:2006rx} where a contracting universe tunnels into an expanding one.\\

Some authors have considered the possibility that the matter collapsing inside a black hole could bounce out in a different universe\!, namely with a different future asymptotic region. The scenario we are considering is  simpler and more conservative: the white hole ``fireworks" emerging from the bouncing black hole takes place where  the black hole is. 
The crucial point demonstrated in \cite{Rovelli2014-2} is that such a simple minimal evolution is {\it possible}: the white hole horizon can be in the future of the black hole horizon, bounding the same external Schwarzschild region with nothing dramatic happening in the surrounding universe.  This unexpected possibility is obtained by carving out the relevant solution from a double covering of the Kruskal metric (where the black hole horizon is in the future of the white hole horizon). This scenario opens the possibility that signals from exploding black holes can be detected and keeps the number of hypotheses at its minimum.

\begin{figure}[t]
\centerline{\includegraphics[width=6cm]{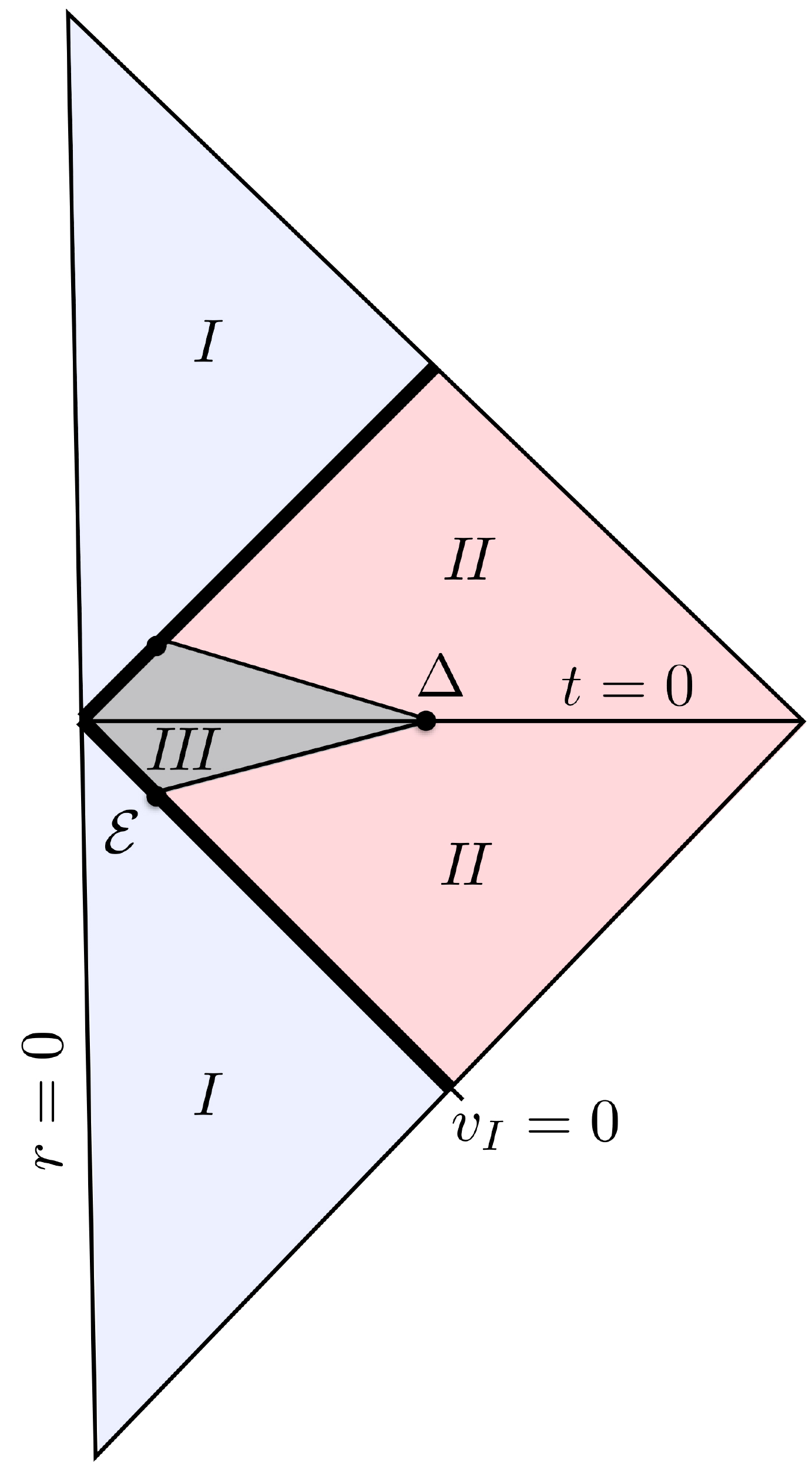}}
\caption{Penrose diagram of a bouncing black hole, from \cite{Rovelli2014-2}.}
\label{ps3}
\end{figure}

The metric found in \cite{Rovelli2014-2} is indeed locally isometric to the Kruskal solution (outside the quantum region), but it is, actually, a portion of a double cover of the Kruskal solution. Fig.\ \ref{ps3} represents a bouncing star where the ``$t=0$" hyperplane is the surface of reflection of the time reversal symmetry in the simplified case of a collapsing null shell. There are two important points (detailed definitions are given in \cite{Rovelli2014-2}), $\Delta$ and $\cal{E}$: the point $\Delta$ at $t=0$ is the maximal extension in space of the region where the Einstein field equations are violated, whereas point $\cal{E}$ is the first moment in time where this happens. Region (I), inside the bouncing shell, must be flat by Bhirkoff's theorem. Region (II), again by Bhirkoff's theorem, must be a portion of the maximal extension of the Schwarzschild metric for a mass $M$. Region (III) is where quantum gravity becomes non-negligible.\\
\indent Because of spherical symmetry, one can use coordinates $(u,v,\theta,\phi)$ with $u$ and $v$ null coordinates in the $r$-$t$ plane and the metric is determined by two functions:
\be
ds^2=-F(u,v) du dv + r^2(u,v)(d\theta^2+\sin^2\theta d\phi^2).
\ee
In these coordinates (in this case called Kruskal--Szekeres coordinates), the Kruskal metric is obtained by taking
\be
F(u,v)=\frac{32M^3}{r}e^{\frac{r}{2m}},
\ee
with $r$ the function of $(u,v)$ defined by 
\be
 \left(1-\frac{r}{2m}\right)e^{\frac{r}{2M}}=uv.
 \ee
The region of interest is bounded by a constant $v={v}_o$ null line drawn thicker in Figure \ref{ps3}.
This constant is a fundamental parameter of the metric under consideration.
The key result of  \cite{Rovelli2014-2} is that, by gluing the different parts of the effective metric, and computing the minimal time for quantum gravitational effects to pile up in the region outside the horizon, one obtains an estimate for the bounce duration: 
 \be
	\tau=-8m \ln v_o> \tau_{q} = 4p\, M^2,
\ee
where $p$ was estimated in \cite{Rovelli2014-2} to be of the order of 0.05. We use Planck units where $G=\hbar=c=1$. The bounce time is therefore proportional to $M^2$, whereas the time of the Hawking evaporation is proportional to $M^3$. Let us write the actual bounce time as $x\tau_{q}$ with $x>1$. This leads to 
\begin{equation}
\tau=4k M^2
\end{equation} 
with $k\equiv xp>0.05$.
As the Hawking evaporation is assumed (in this model) to be a small dissipative process that can be neglected at first order, $k$ cannot be taken arbitrarily large. If $k$ is too large the bouncing time becomes comparable to the Hawking time and the model fails. Therefore, there exists a given interval (which, in principle, depends on $M$) of values of $k$ for which the scenario is consistent. Unfortunately, this interval is very large. In this article we investigate all possibilities by considering the whole range of the possible values of $k$, yielding different characteristics for the observable signals.

The basic phenomenology was investigated in \cite{Barrau2014-2} where $k$ was assumed to take its smallest possible value, that corresponds to the shortest bounce. The aim of the present article is to go beyond this first study, following in two directions. First, we generalize the previous results by varying $k$. 
The assumption that the bounce time remains smaller than the Hawking time is supported by the ``firewall argument'' presented in \cite{Rovelli2014}, in the sense that it allows to solve the information paradox without requiring something particular to happen at the horizon (the equivalence principle is therefore respected). We study in detail the maximal distance at which a single black-hole bounce can be detected. Second, we go beyond the ``single event detection" and consider the diffuse emission produced by a distribution of bouncing black holes on cosmological scales. 

\section{Single event detection}

For detection purposes, we are interested in black holes with a lifetime that is less than the age of our universe.
Therefore, for a PBH detected today, this condition translates into $\tau=t_H$ where $t_H$ is the Hubble time. This fixes the mass $M$, as a function of the parameter $k$ (defined in the previous section). In all the considered cases, $M$ remains very small compared to a solar mass and would correspond to PBHs possibly formed in the early Universe. Although no PBH has been detected to date, various mechanisms for their production shortly after the Big Bang have been suggested (see, {\it e.g.}, \cite{Carr1975} for an early detailed calculation and \cite{green} for a review). Although their number density might be way too small for direct detection, the production of PBHs remains a quite generic prediction of cosmological physics either directly from density perturbations --possibly enhanced by phase transitions-- or through exotic phenomena like the collapse of string loops generated by string self-intersections or collisions of bubbles of false vacua.\\
\indent The energy (and amplitude) of the  signal emitted in the quantum gravity model considered here remains an open issue. As suggested in \cite{Barrau2014-2} and to remain general, we consider two possible signals of different origins. The first one, referred to as the {\it low energy} signal, is determined by dimensional arguments.  The white hole horizon from which matter emerges has size $L\approx 2M$ and its emission, in the metric studied in \cite{Rovelli2014-2}, is instantaneous. It is natural to expect that the signal of an exploding object includes a component with a wavelength equal to its size.  This is the main scale of the problem and it fixes an expected wavelength for the emitted radiation: $\lambda \approx L$. 
At this stage no detailed astrophysical model is available for this component. It should be noticed that although the instantaneous Hawking radiation is also emitted with a wavelength of order $L\approx 2M$, the Hawking evaporation is a quasi-continuous process while the phenomenon we are considering here is a tunneling-like phenomenon: a sudden explosion where the entire energy of the hole is emitted together.  The two phenomena are therefore very different when considering the time integrated spectra.  The signal we are considering, in particular, is different from what was investigated in \cite{cline}, except for some particular values of $k$. This has been studied in \cite{Barrau2014}. We assume that particles are emitted at the prorata of their number of internal degrees of freedom. (This is also the case for the Hawking spectrum in the optical limit, {\it i.e.} when the greybody factors describing the backscattering probability are spin-independent.)\\
\indent The second signal, referred to as the {\it high energy} component, has a very different origin. Consider the history of the matter emerging from a white hole: it comes from the bounce of the matter that formed the black hole by collapsing. In most scenarios there is a direct relation between the initial mass $M$ of a PBH and the temperature of the Universe when it was formed (see \cite{carr_rev} for a review). The black hole mass, $M$, should be of the order of the horizon mass\footnote{Other more exotic models, {\it e.g.} collisions of cosmic strings or collisions of bubbles associated with different vacua, can lead to different masses at a given cosmic time. We will not consider them in this study.}, $M_H$:
\begin{equation}
M\sim M_H\sim t .
\end{equation}
The cosmic time $t$ is related to the temperature of the Universe $T$ by
\begin{equation}
 t\approx 0.3g_*^{-\frac{1}{2}}\, T^{-2}, 
 \end{equation}
where $g_*\approx 100$ is the number of degrees of freedom. Once $k$ is fixed, $M$ is fixed (by $\tau\approx t_H$) and  $T$ is therefore known. As the process is time-symmetric, what comes out from the white hole should be what went in the black hole, re-emerging at the same energy: a blackbody spectrum at temperature $T$. Intuitively, the bouncing black hole plays the role of a ``time machine" that sends the primordial universe radiation to the future: while the surrounding space has cooled down to 2.7K, the high-energy radiation emerges from the white hole with its original energy.\\
\indent When the parameter $k$ is taken larger than its smallest possible value, which is fixed by the requirement that quantum effects are important enough to lead to a bounce, the bounce time becomes larger for a given mass. If this time is assumed to be equal to the Hubble time (or slightly less if we focus on black holes bouncing far away), this means that the mass has to be smaller. The resulting energy will then be higher for both the {\it low energy} and the {\it high energy} signals, but for different reasons. In the first case, this is because of the smaller size of the hole, leading to a smaller emitted wavelength. In the second case, the reason is a bit more subtle: the  primordial black hole has to be formed earlier, when the Hubble mass was smaller, and the temperature of the Universe was therefore higher. Importantly, we will show later that although both signals vary in the same ``direction" as a function of $k$, they do not have the same $k$-dependence. \\
\begin{figure}[t!]
\centerline{\includegraphics[width=9cm]{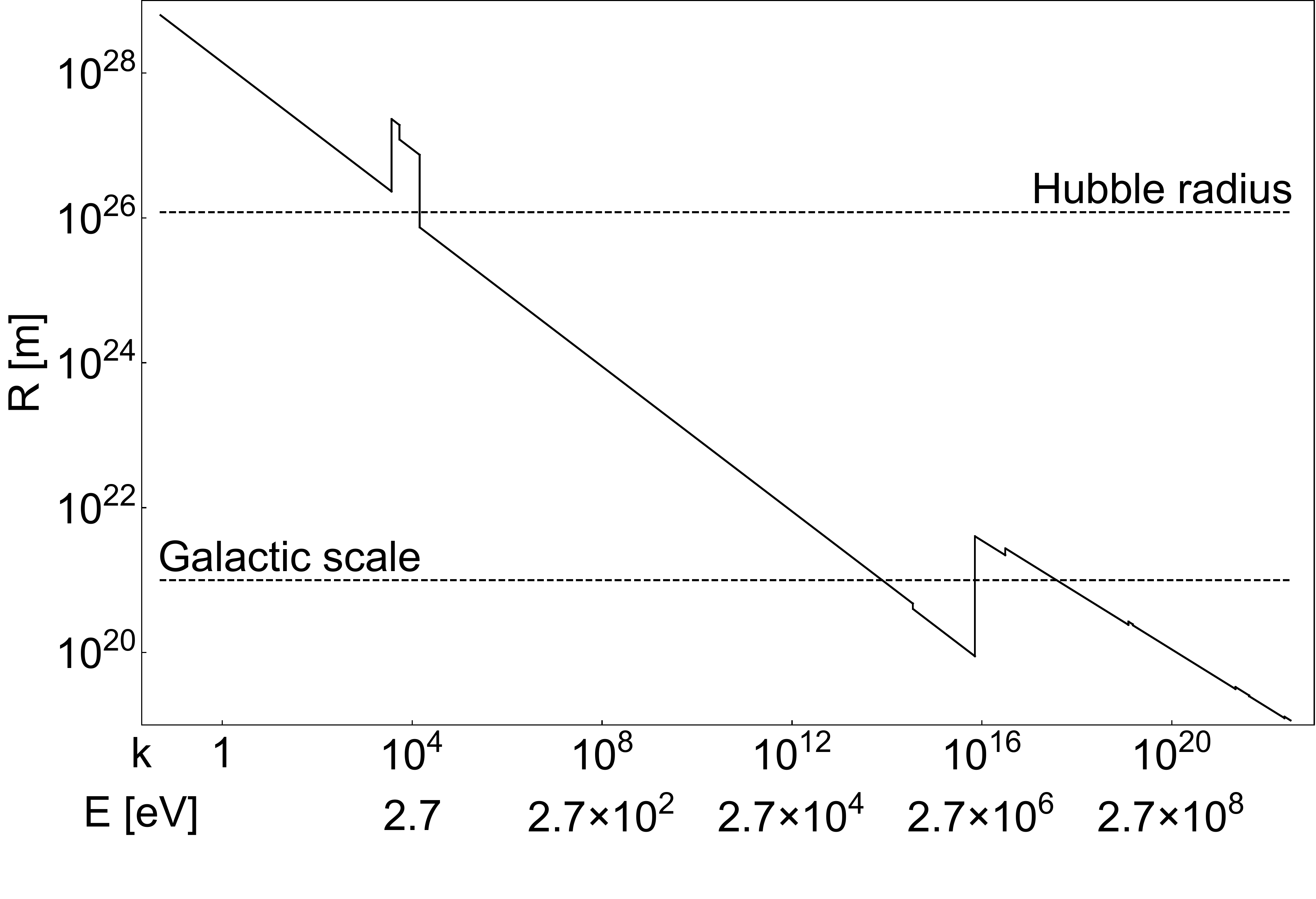}}
\caption{Maximum distance at which a bouncing black hole can be detected in the  {\it low energy} channel as a function of the parameter $k$ (the associated signal energy is also given). The upper horizontal dashed line represents the Hubble radius and the lower one represents the Galactic scale.}
\label{distlow}
\end{figure}
The first question to be addressed is the  maximal distance at which one could observe a bouncing black hole. We focus on emitted photons as, unlike charged cosmic-rays, they travel in straight lines and therefore allow for a precise determination of the event location. When $k$ varies from its minimum value ($\approx 0.05$, determined for the quantum effects to cumulate and the bounce to take place) to its maximum value ($\approx 10^{22}$, determined for the bounce time to remain smaller than Hawking time), the energy of the emitted signal varies over many orders of magnitude.\\

The resulting detectability depends on several factors:
\begin{itemize}
\item {\it The size of the detector (and its detection efficiency).} In the infrared, ultra-violet, X-rays and soft gamma-rays, only satellites can be used as the atmosphere is not transparent. This fixes the size around an order of magnitude close to a meter. In the optical domain, larger ground-based telescopes are available (around ten meters). For hard gamma-rays the size of the instrument is no longer relevant, what matters is the size of the Cherenkov shower induced by the high-energy photon. This increases the size to roughly a hundred meters.
\item {\it The absorption during the  propagation over cosmological distances.} Although some subtleties do appear at several energies, the Universe is mostly transparent up to TeV energies where pair production of leptons becomes possible through interactions with the cosmic infrared background $\gamma_{TeV}+\gamma_{IR}\rightarrow e^++e^-$. The absorption is basically exponential above the threshold energy (corresponding to twice the electron mass in the center-of-mass frame of the interaction).
\item {\it The number of measured photons required for the detection to be statistically significant,} that is to be several standard deviations above the background fluctuations. This is also energy-dependent. For example, although a few synchronous measured gamma-rays are enough for a detection in the 100 GeV range --where the background is very low-- many more are required in the optical band. Not only because the diffuse background is much higher at lower energies, but also because measurements require a substantial integration time that makes the determination of the accurate arrival timing impossible. We have only used crude approximations for the galactic and extragalactic backgrounds assuming that the line of sight aways lies outside the galactic plane. A more careful analysis would be important in the future.
\end{itemize}

Figure \ref{distlow} represents the maximum distance at which a bouncing black hole can be seen in the {\it low energy} channel, calculated by taking into account all the above-mentioned phenomena. Several effects of different origins can be observed. The large step around $k=10^4$ is associated with the larger size of ground based optical telescopes. The little steps decreasing the distance are associated with the fact that the mean energy of the signal emitted by the bouncing black hole becomes higher than the mass of a new particle: this new particle can then be emitted and if it does not decay into photons, the percentage of produced photons inevitably decreases (and so does the maximum distance). 
\begin{figure}[t]
\centerline{\includegraphics[width=9cm]{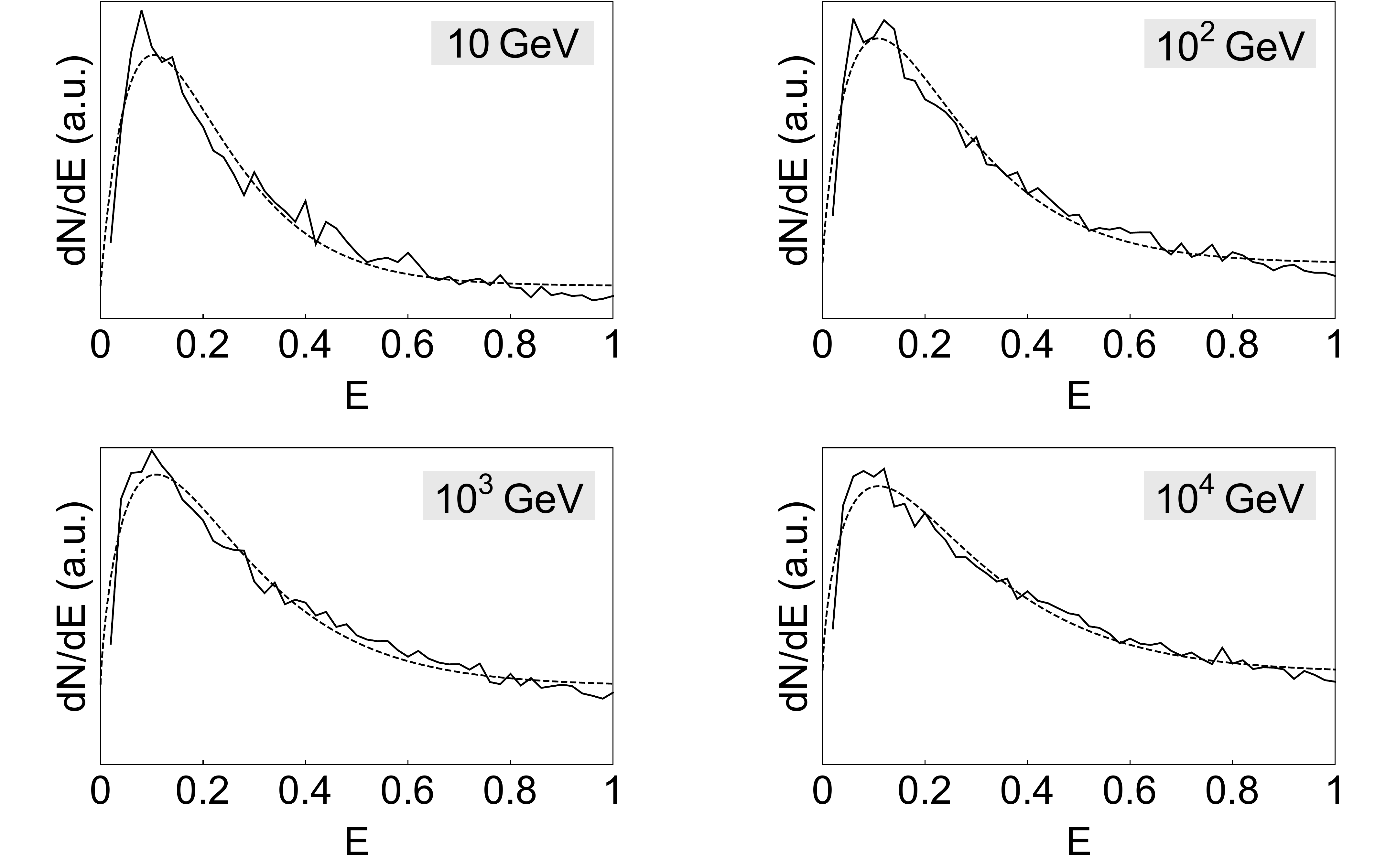}}
\caption{Histograms of gamma-rays produced by jets of quarks at 10, 100, 1000, and 10000 GeV. The smooth curves are the fits used in the following analysis whereas the other curve is the output of the Monte-Carlo simulation with 10000 events. The energies of the $x$ axis are given in GeV.}
\label{had}
\end{figure}
An important step in the opposite direction occurs around $k=10^{16}$. This is due to the fact that quarks begin to be emitted. Then, the most important source of gamma-rays emitted by the bouncing black hole is no longer the direct emission --that is photons emitted as such by the PBH-- but instead the one coming from the decay of neutral pions (whose lifetime is negligible here) produced in the fragmentation process of the emitted partons. To take this into account, we have used the Pythia program  \cite{tor}, which is a standard tool for the generation of events in high-energy collisions, comprising a coherent set of physics models for the evolution from a few-body hard process to complex multiparticle final states. It incorporates a large number of hard processes, models for initial and final state parton showers, matching and merging methods between hard processes and parton showers, multiparton interactions, beam remnants, string fragmentation and particle decays. It is based on the Lund model \cite{lund}. Although most previous approaches have used cruder analytical approximations, this way of treating the quark and gluon emission is not new and was also implemented in the study of hadron production by primordial black holes: as soon as the black hole temperature becomes higher than the quantum chromodynamics (QCD) confinement scale, those processes inevitably have to be taken into account \cite{macgibbon}. In a high-energy hadronic process, a very large number of pions can be generated. As nearly each neutral pion will decay into two photons, this mechanism --called ``indirect" or ``secondary" emission-- will, by far, dominate the gamma-ray production. In Fig. \ref{distlow}, little steps increasing the maximum distance can also be seen. They are due to the fact that the available energy reaches a new threshold corresponding to the possible emission of  a new quark --because the black hole size becomes smaller than its inverse mass-- that will produce gamma-rays in its hadronization. This leads to more gamma-rays (whereas at lower energies, or lower values of $k$, the emission of new particles was only associated with a lower gamma-ray rate). \\

Figure \ref{had} shows the histograms obtained using Pythia for different jet energies. The smooth curves corresponds to the fits used in the analysis. It is interesting to note that increasing the available energy increases the number of generated gamma-rays and the mean energy of the histogram but not the position of the peak in the distribution which is associated with $\pi^0$ particles generated at rest in the galactic frame.

\begin{figure}[h!]
\centerline{\includegraphics[width=9cm]{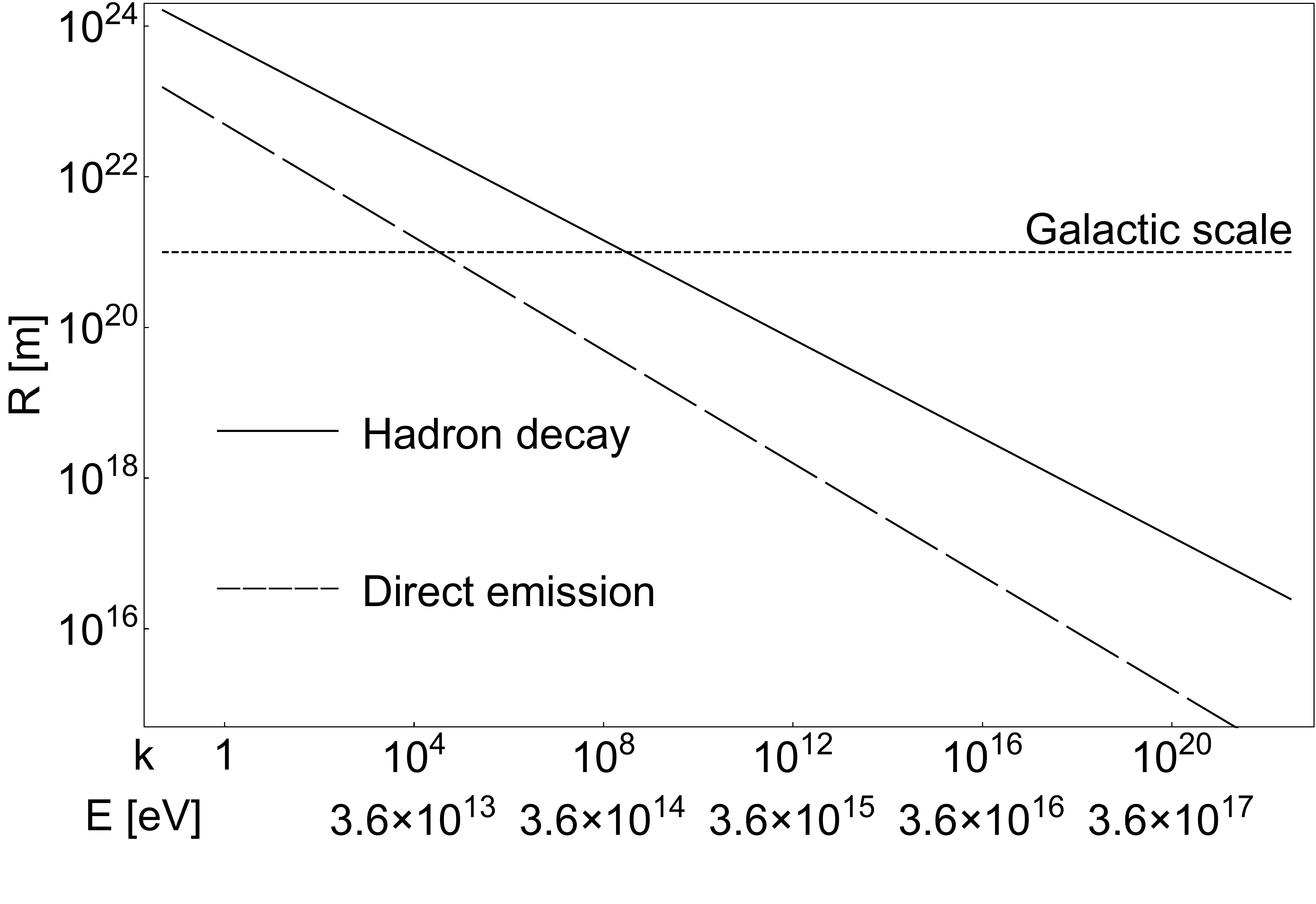}}
\caption{Maximum distance at which a bouncing black hole can be detected in the {\it high energy} channel as a function of the parameter $k$ (the associated signal energy is also given). The horizontal dashed line represents the Galactic scale. The lower line corresponds to the direct emission and the upper one to the decay of unstable hadrons produced by jets of quarks and gluons. Interestingly, the slope is not exactly the same for both signals.}
\label{disthigh}
\end{figure}

Figure \ref{disthigh} represents the maximum distance at which a bouncing black hole can be seen in the {\it high energy} channel. The lower curve represents the direct emission of gamma-rays and the higher one represents gamma-rays coming from the decay of unstable hadrons. As expected, the latter dominates. For this signal, there is no threshold effect associated with masses as the effective temperature of the process is in any case well above the QCD confinement scale.\\

The largest distance is given, for any $k$, by $d=inf \{d_{hor}, sup\{d_{direct},d_{decay}\}\}$ where $d_{decay}$ is the maximum distance for the photons associated with the secondary emission while $d_{direct}$ is the one associated with the direct emission and $d_{hor}$ is the horizon 
at the considered energy. 
Photons cannot come from arbitrary long distance and are limited by an effective horizon
ranging from around a Gpc for photons in the 100 GeV - 1 TeV range to a few Mpc at 100 TeV because of their interactions with the diffuse background. 
Above this energy, interactions with the CMB become possible  and the horizon can decrease to a few kpc only around 1000 TeV. This effect does not happen for the indirect emission which takes place at a lower energy where the Universe is quite transparent. Although the effective surface of detectors (due to Cherenkov showers) is much higher at high energy this does not compensate for the limited flux. The flux is small at high energy for two reasons. First, because it is associated with smaller PBH masses, making the total energy available smaller. Second, because the energy carried out by each emitted photon is higher, making their number smaller even for the same total available energy.

\indent It is interesting to investigate  analytically the $k$ dependence of both signals. In both cases, one can use the following approximation for the relationship between time and redshift:
\begin{equation}t\approx \frac{2H_0^{-1}}{3\Omega_{\Lambda}^{1/2}}{\rm sinh}^{-1}\left[\left(\frac{\Omega_{\Lambda}}{\Omega_{M}}\right)^{\frac{1}{2}}(1+z)^{-\frac{3}{2}}\right],
\end{equation}
where $H_0$ is the Hubble constant, $\Omega_{\Lambda}$ is the normalized dark energy density, and $\Omega_{M}$ is the normalized matter density. Requiring this time to be equal to the bounce time $4kM^2$  leads, for the measured {\it low energy} signal, to
\begin{equation}\lambda_{low}^{meas}\approx
2\,(1+z)\sqrt{\frac{H_0^{-1}}{6k\Omega_{\Lambda}^{1/2}}{\rm sinh}^{-1}\left[\left(\frac{\Omega_{\Lambda}}{\Omega_{M}}\right)^{\frac{1}{2}}(1+z)^{-\frac{3}{2}}\right]}.
\end{equation}\\
\indent The same reasoning can be applied to the {\it high energy} signal. To fix orders of magnitude, one can write 
$\lambda\approx \frac{2\pi}{k_BT}$ where $k_B$ is the Boltzmann constant and $T$ is the temperature of the Universe at the formation time. Gathering everything, this leads to
\begin{widetext}
\begin{equation}\lambda_{high}^{meas}\approx
\frac{2\pi}{k_BT}\frac{(1+z)}{(0.3g_*^{-1})^{\frac{1}{2}}}\left[\frac{H_0^{-1}}{6k\Omega_{\Lambda}^{1/2}}{\rm sinh}^{-1}\left[\left(\frac{\Omega_{\Lambda}}{\Omega_{M}}\right)^{\frac{1}{2}}(1+z)^{-\frac{3}{2}}\right]\right]^{\frac{1}{4}}.
\end{equation}
\end{widetext}
Although the mean wavelength does decrease as a function of $k$ in both cases, it does not follow the same general behavior. It scales with $k^{-\frac{1}{2}}$ for the low energy component and as $k^{-\frac{1}{4}}$ for the high energy one.The following conclusions can be drawn:
\begin{itemize}
\item The {\it low energy} channel leads to a better single-event detection than the {\it high energy} channel. Although a lower energy dilutes the signal in a higher astrophysical background, this effect is over-compensated by the larger amount of photons.
\item The difference of maximal distances between the {\it low-} and {\it high energy} channels decreases for higher values of $k$, {\it i.e.} for longer black-hole lifetimes.
\item In the {\it low energy} channel, for the smaller values of $k$, a single bounce can be detected arbitrary far away in the Universe.
\item In all cases, the distances are large enough and experimental detection is far from being hopeless.
\end{itemize}

\section{Integrated emission}

In addition to the instantaneous spectrum emitted by a single bouncing black hole, it is interesting to consider the possible diffuse background due to the integrated emission of a population of bouncing black holes. Formally, the number of measured photons detected per unit time, unit energy and unit surface, can be written as:
\begin{equation}
\frac{dN_{mes}}{dEdtdS}=\int\Phi_{ind}((1+z)E,R) \cdot n(R) \cdot A(E) \cdot f(E,R)  dR,
\label{flux_int}
\end{equation}
where $\Phi_{ind}(E,R)$ denotes the individual flux emitted by a single bouncing black hole at distance $R$ and at energy $E$, $A(E)$ is the angular acceptance of the detector multiplied by its efficiency, $f(E,R)$ is the absorption function, and $n(R)$ is the number of black holes bouncing at distance $R$ per unit time and volume. The distance $R$ and the redshift $z$ entering the above formula are linked. The integration has to be carried out up to cosmological distances and it is therefore necessary to use exact results behind the linear approximation. The energy is also correlated with $R$ as the distance fixes the bounce time of the black hole which, subsequently, fixes the emitted energy.\\

It is worth considering  the $n(R)$ term in more details. If one denotes by $\frac{dn}{dMdV}$ the initial differential mass spectrum of primordial black holes per unit volume, it is possible to define $n(R)$ as:
\begin{equation}
n(R)=\int_{M(t)}^{M(t+\Delta t)}\frac{dn}{dMdV}dM,
\end{equation}
leading to
\begin{equation}
n(R)\approx\frac{dn}{dMdV}\frac{\Delta t}{8k}, 
\end{equation}
where the mass spectrum is evaluated for the mass corresponding to a time $(t_H-\frac{R}{c})$. The shape of the mass spectrum obviously depends on the details of the formation mechanism (see \cite{carr1994} for a review on PBHs and inflation). As an example, we shall assume that primordial black holes are directly formed by the collapse of density fluctuations with a high-enough density contrast in the early Universe. The initial mass spectrum is then directly related to the equation of state of the Universe at the formation epoch. It is given by \cite{jane, Carr1975}: 
\be
\frac{dn}{dMdV}=\alpha M^{-1-\frac{1+3w}{1+w}},
\label{spec}
\ee
where $w=p/\rho$. In a matter-dominated universe the exponent $\beta\equiv-1-\frac{1+3w}{1+w}$ takes the value $\beta=-5/2$. The normalization coefficient $\alpha$ will be kept unknown as it depends on the details of the black hole formation mechanism. For a sizeable amount of primordial black holes to form, the power spectrum  normalized on the CMB needs to be boosted at small scales. The formula given above might therefore be correct only within a limited interval of masses. The idea is that the mass spectrum takes a high enough value in the relevant range whereas it is naturally suppressed at small masses by inflation. We will neither study those questions here (focusing on the shape of the resulting emission), nor the normalisation issues which depend sensitively on the bounds of the mass spectrum, that are highly model-dependent. As this  part of the study is devoted to the investigation of the shape of the signal, the $y$ axis on the figures are not normalized. As we show below, the shape of the signal is quite independent on the shape of the mass spectrum, so Eq.\,\ref{spec} does not play any significant role for the computed spectra. \\

\begin{figure}
\centerline{\includegraphics[width=8.5cm]{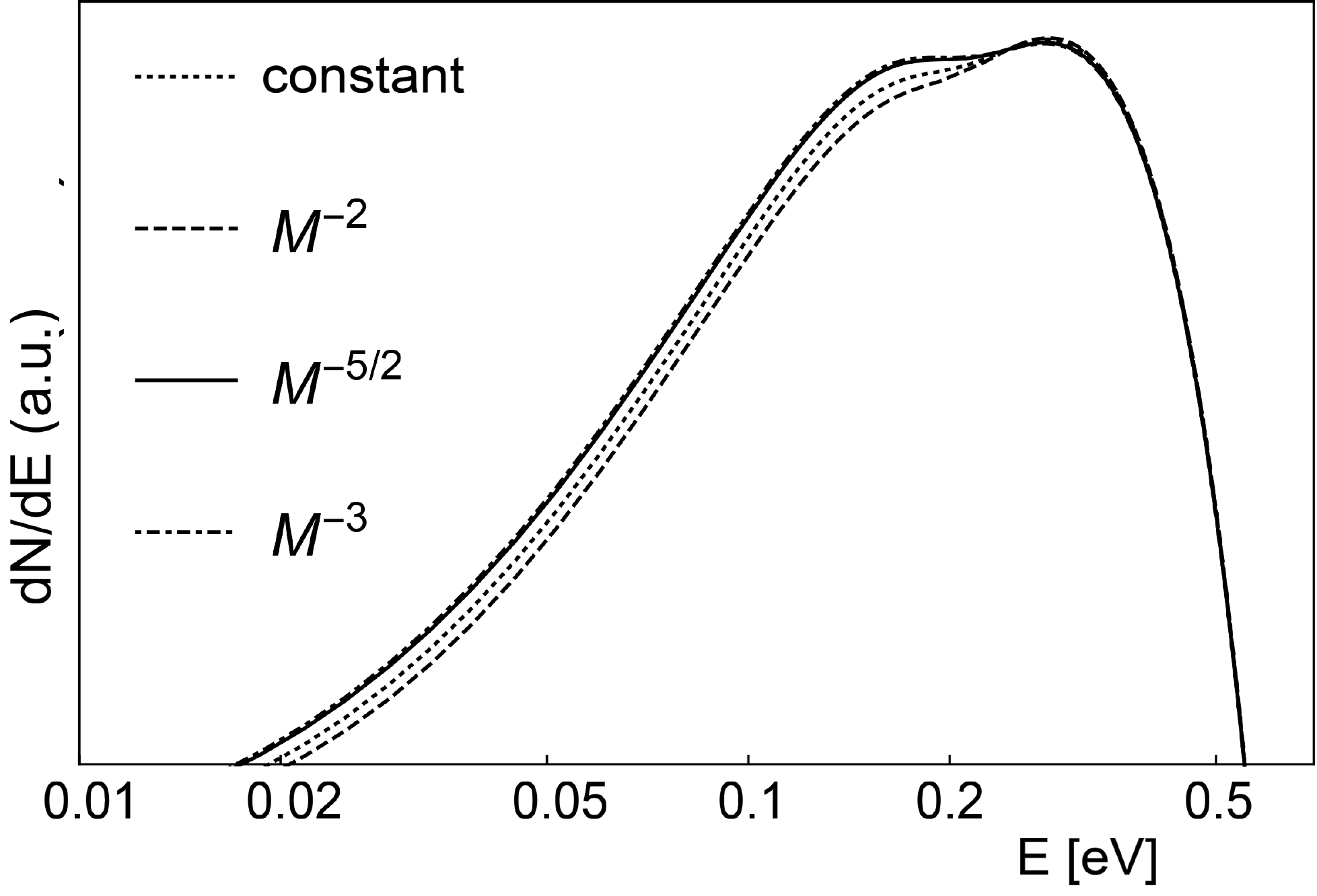}}
\caption{{\it Low energy} channel signal calculated for different mass spectra. As the mass spectrum is not normalized, the units of the $y$ axis are arbitrary.}
\label{mass_spectr}
\end{figure}
\begin{figure}
\centerline{\includegraphics[width=9cm]{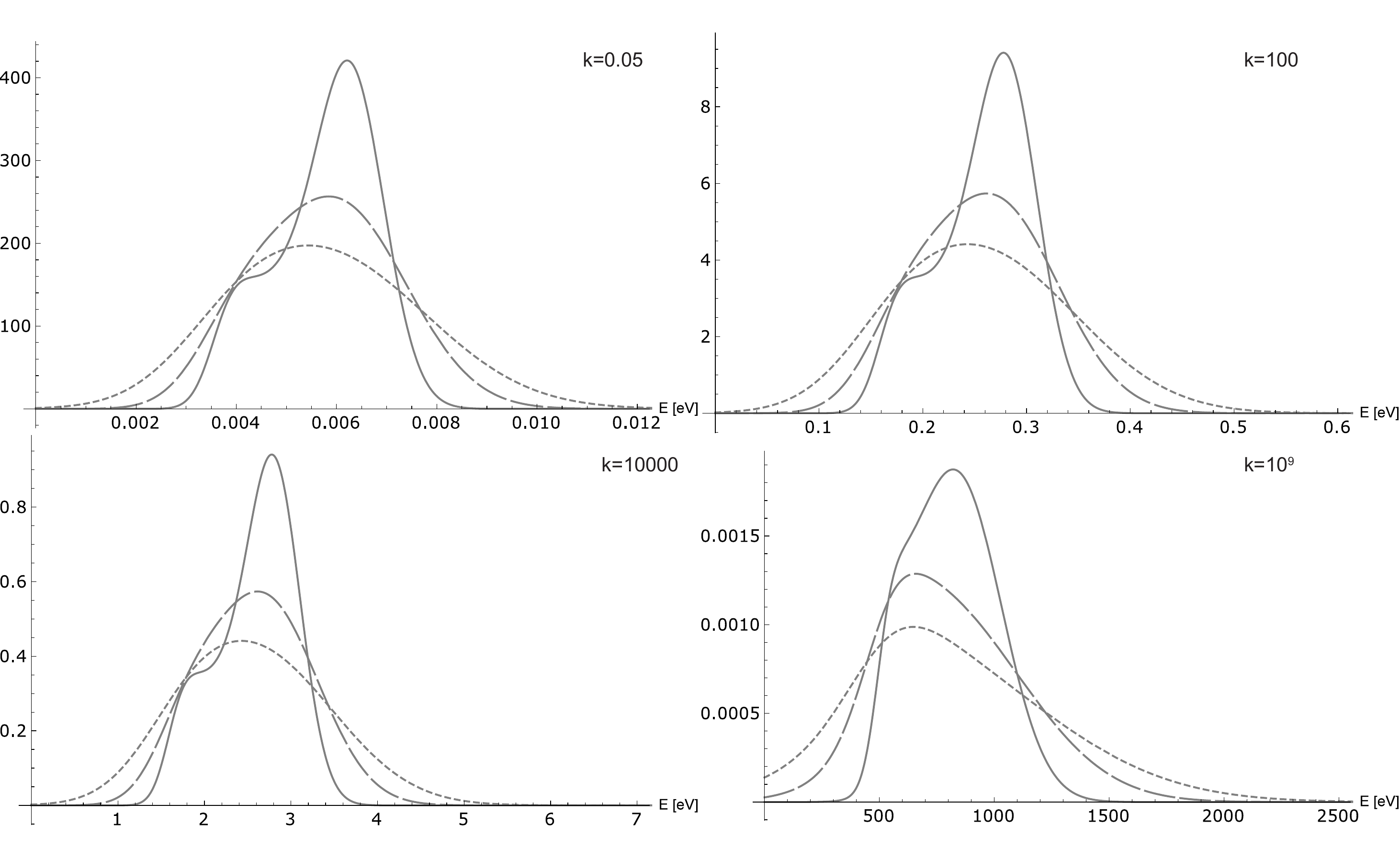}}
\centerline{\includegraphics[width=9cm]{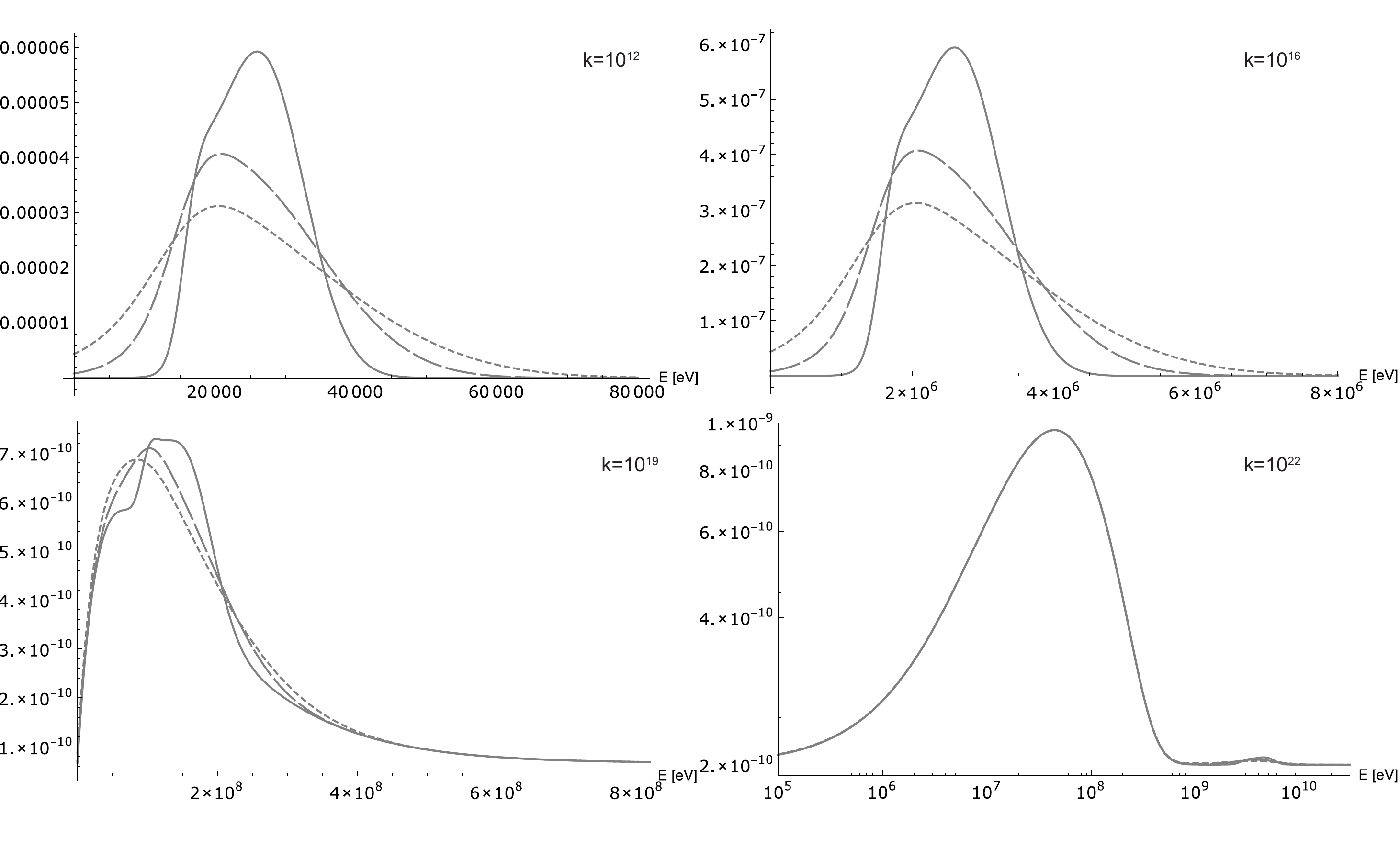}}
\caption{Electromagnetic signal expected, in the {\it low-energy} channel, from a distribution of bouncing black holes respectively with $k=0.05,\, k=100,\, k=10000,\, k=10^5,\, k=10^{12},\, k=10^{16},\, k=10^{19},\, k=10^{22}$. The plain line corresponds to $\sigma=0.1E$, the dashed line to $\sigma=0.2E$ and the dotted line to $\sigma=0.3E$, where $E$ is the mean energy of emitted signal. As the mass spectrum is not normalized, the units of the $y$ axis are arbitrary.}
\label{integ_low}
\end{figure}

This is illustrated in Fig. \ref{mass_spectr} where different hypothesis for the exponent $\beta$ are displayed. The  electromagnetic spectrum induced by the distribution of bouncing black holes is almost exactly the same. Only one case ($\beta=-5/2$, corresponding to $w=1/3$) is therefore considered in the following, leading to generic results.
The black holes are assumed to be uniformly distributed in the Universe, which is a meaningful hypothesis as long as we deal with cosmological distances\footnote{
The local distribution of primordial black hole is expected to match dark-matter distribution.
}.\\

Once again, we consider the two different channels for the emitted signal. Let us begin with the {\it low energy} signal. The issues about the different components of the emitted photons, presented in the preceding section, are still accounted for in this part. Figure \ref{integ_low} displays the resulting signal on Earth for values of the parameter $k$ varying from 0.05 (minimum) to $10^{22}$ (maximum) by carrying out the numerical integration of Eq. \ref{flux_int}. The smallest value of $k$ minimizes the bounce time whereas the largest one makes it comparable to the Hawking time. The last plot of Fig. \ref{integ_low} is in double logarithmic scale to improve the readability. On this plot, it is easy to distinguish the direct emission (on the right side) from the emission due to the decay of pions produced by the fragmentation of parton jets (on the left side). Obviously, the second strongly dominates. For smaller values of $k$, there is a little ``bump" in the signal which is due to the non-linear redshift-distance relation leading to a kind of ``accumulation" of the signal. In principle, by construction of the model, the direct emission is nearly monochromatic. This is however obviously an approximation and we have therefore considered three possible relative widths for the signal : $\sigma/E=0.1$, $\sigma/E=0.2$ and $\sigma/E=0.3$ where $E$ is the mean energy of emitted quanta. Those three hypotheses are displayed on the plots of Fig. \ref{integ_low}.\\

\begin{figure}[t]
\centerline{\includegraphics[width=9cm]{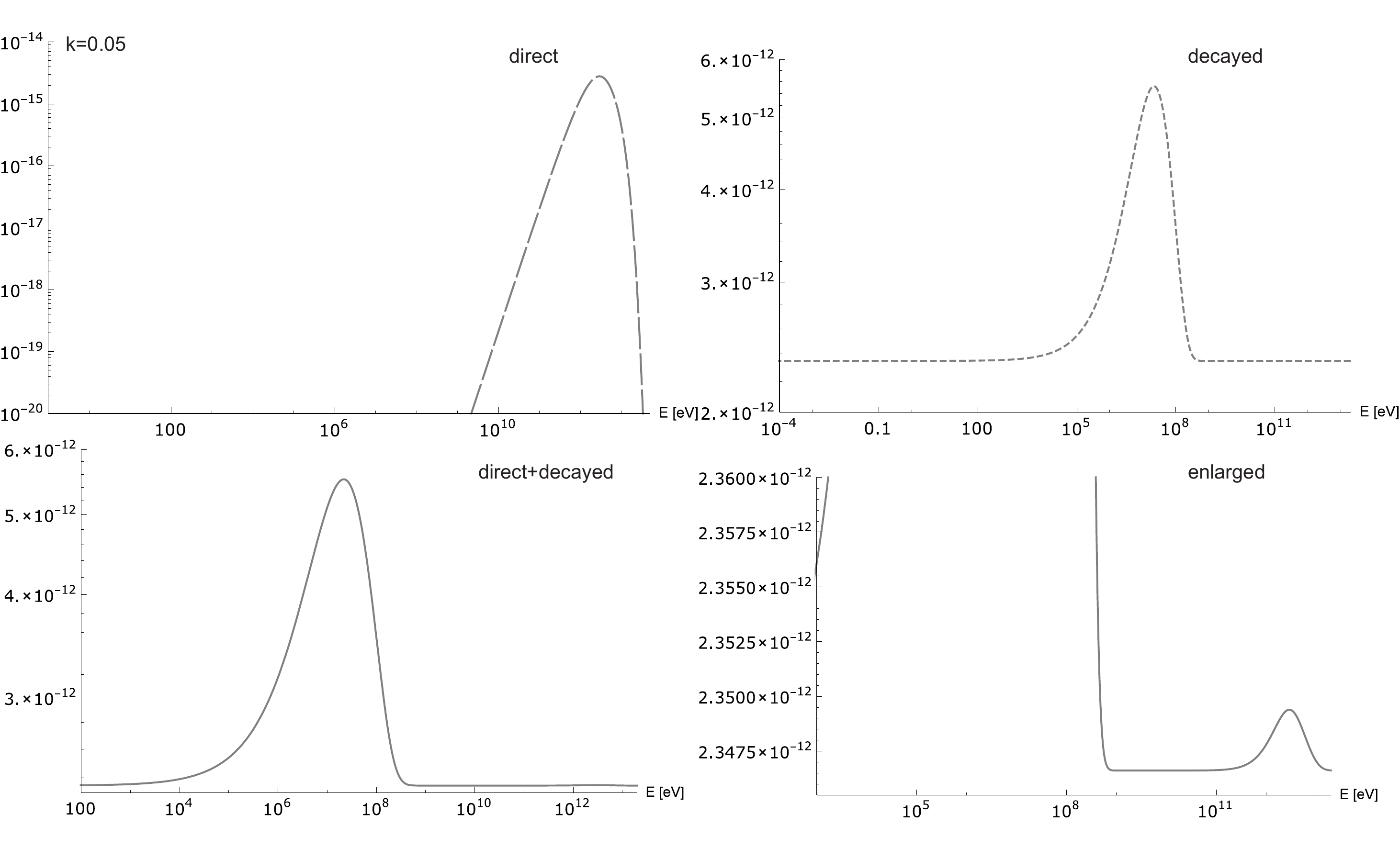}}
\caption{Electromagnetic signal expected, in the {\it high-energy} channel, from a distribution of bouncing black holes with $k=0.05$. The upper left plot corresponds to the direct emission, the upper right plot to the gamma-rays coming from the decay of neutral pions produced by jets of quarks. The lower left plot is the full signal. The lower right plot is a zoom on the direct emission part of the spectrum.}
\label{integ_high1}
\end{figure}

\begin{figure}[t]
\centerline{\includegraphics[width=9cm]{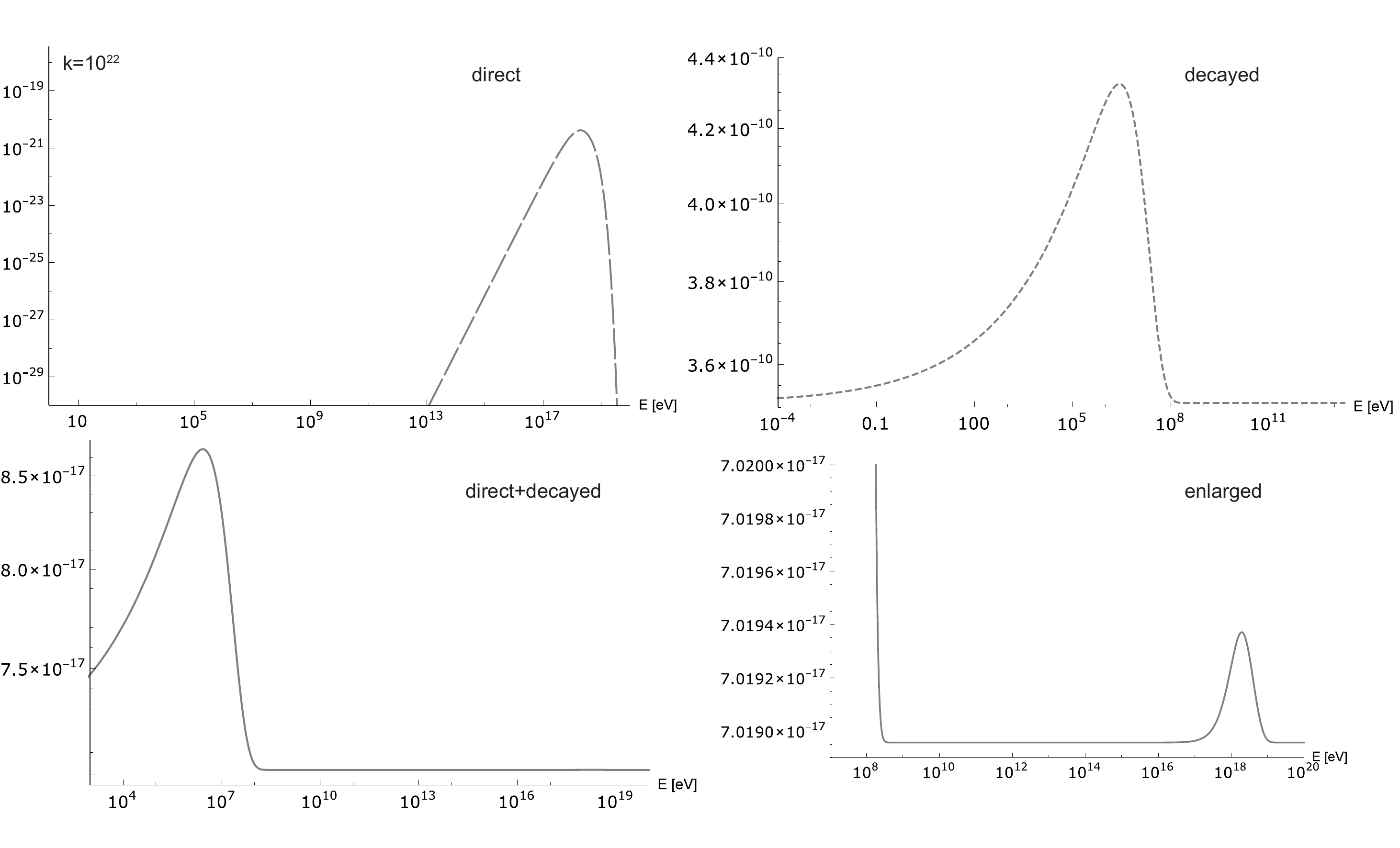}}
\caption{Electromagnetic signal expected, in the {\it high-energy} channel, from a distribution of bouncing black holes with $k=10^{22}$. The upper left plot corresponds to the direct emission, the upper right plot to gamma-rays coming from the decay of neutral pions produced by jets of quarks. The lower left plot is the full signal. The lower right plot is a zoom on the direct emission part of the spectrum.}
\label{integ_high2}
\end{figure}

We have also considered the {\it high energy} signal coming from radiation re-emitted at the energy at which it was absorbed in the early universe. In this case, the situation is better controlled as the spectrum of the signal is accurately known: it is simply given by a blackbody law at the temperature $T$ of the plasma filling the Universe at the formation time of the black hole. As the shape of the signal is very weakly dependent on the value of $k$, we have just displayed the two extreme cases ($k=0.05$ and $k=10^{22}$) in Fig. \ref{integ_high1} and Fig. \ref{integ_high2}. As it can be seen on the plots, the emission is strongly dominated by the gamma-rays coming from decaying neutral pions.\\

For both the {\it low energy} and the {\it high energy} signals the integration effect does not change much the signal as it would be expected from a single bouncing black hole. This is due to a ``redshift-compensation" effect. When considering a black hole bouncing far away, the mean energy of its emitted signal (in its rest frame) is smaller for both the {\it low energy} and the {\it high energy} cases but, as explained before, for different reasons. In the first case, this is because a black hole observed now and bouncing far away has a smaller lifetime, so that  its initial mass is therefore smaller and so is its radius. As the emission wavelength is controlled by the size of the black hole, the emitted signal has a higher energy. It should also be underlined that the more distant the black hole, the smaller the number of emitted photons. This is not only because the total available energy (given by the mass of he black hole) is smaller but also because the individual energy of each photon is, in addition, higher. For the second case, the {\it high energy} emission, a black hole bouncing far away also has a smaller lifetime, hence a smaller initial mass: it was formed earlier in the history of the Universe 
(in the ``standard" formation scenario we are considering), when the plasma was hotter. The emitted signal (which is the same as the absorbed one in this case) is higher in energy as well.  In both cases, this higher emitted energy is partially compensated by the redshift, therefore reducing the distortion induced by the integration effect. \\

The conclusion is that the shape of the signal might be used as an observational signature of its specific origin in the {\it high energy} case. It looks indeed like a slightly distorted (by the redshift-distance integration) blackbody law that is not to be expected from any other known astrophysical effect. In the {\it low energy} case, the situation is less clear as the accurate shape (in particular width) on the signal is unknown but, still, quite generic features do appear in the figures, leading to some hope for detection.\\

It is still impossible to normalize these plots so as to compare them with the astrophysical background, whose spectral energy density roughly scales as $E^{-1/2}$, as this entirely depends on the percentage of dark matter made by PBHs and, more importantly, on the arbitrary choice of the bounds on the mass spectrum.

\section{Conclusion}

The possibility that black holes are bouncing objects, suggested by quantum-gravity arguments, should be taken seriously. We have studied the individual bounce detectability and the integrated signal for all possible values of the bounce time. Some characteristic features emerge. 

This study should be pushed forward in two directions. On the theoretical side, it would be very interesting to compute the quantum transition amplitudes between the contracting classical black hole solution and the expanding classical white hole solution \cite{Rovelli-talk}. Explicit models of quantum gravity, {\it e.g.} loop quantum gravity, do, in principle, make this calculation possible.

On the phenomenological side, it would be important to consider not only photons but also charged cosmic-rays that should be emitted by bouncing black holes as well. New experimental data are being made available (in particular by the AMS experiment onboard the International Space Station) and any predicted excess could be detectable in the near future. Although the signal looses its directionality and is confined to smaller scales, the enhancement by the galactic magnetic field could lead to promising detection perspectives.
Finally, it would also be interesting to use known constraints on primordial black holes to investigate how they should be revised in this model, for different values of the $k$ parameter.




\end{document}